%% file: main__1_.tex
\documentclass[10pt,twocolumn]{article}

\usepackage[T1]{fontenc}
\usepackage[utf8]{inputenc}
\usepackage{lmodern}
\usepackage[margin=0.72in,columnsep=0.25in]{geometry}
\usepackage{microtype}
\usepackage{graphicx}
\usepackage{booktabs}
\usepackage{tabularx}
\usepackage{enumitem}
\usepackage{xcolor}
\usepackage{balance}
\usepackage[authordate,backend=biber,natbib=true,doi=true,eprint=false,isbn=false,url=true]{biblatex-chicago}
\usepackage{hyperref}

\definecolor{linkblue}{HTML}{174A7E}
\hypersetup{
  colorlinks=true,
  linkcolor=linkblue,
  citecolor=linkblue,
  urlcolor=linkblue,
  pdftitle={Towards Assurance Closure in AI-Native Large-Scale Agile Software Development},
  pdfauthor={Ricardo Britto}
}

\setlist[itemize]{leftmargin=*,nosep,topsep=2pt}
\setlist[enumerate]{leftmargin=*,itemsep=2pt,topsep=2pt}
\title{\vspace{-1.2em}\textbf{Towards Assurance Closure in AI-Native Large-Scale Agile Software Development}}

\author{Ricardo Britto\\
\small Ericsson, Sweden\\
\small Blekinge Institute of Technology, Sweden\\
\small \texttt{ricardo.britto@ericsson.com}}
\date{August 2026}

\begin{document}
\maketitle
\vspace{-1.2em}

\begin{abstract}
The AI-Native Manifesto envisions large-scale agile software development in which humans increasingly govern intent, risk, and exceptions while agents execute more of the engineering process. Realizing that end-state requires more than better code generation: it requires \emph{assurance closure}, meaning that the system can establish what must be true, determine and obtain appropriate evidence, judge the credibility of that evidence, preserve its validity through change, and use the resulting uncertainty to bound agent authority. Existing work already provides many of the necessary mechanisms across formal methods, testing, simulation, assurance cases, digital twins, and runtime assurance. We identify six residual gaps in making the surrounding assurance reasoning sufficiently machine-operable, propose a high-level architecture with six corresponding capabilities built on a shared semantic assurance layer, and formulate four research questions to turn that architecture into dependable, human-on-the-loop, AI-native R\&D.
\end{abstract}

\input{sections/01_position}
\input{sections/02_gaps}
\input{sections/03_architecture}
\input{sections/04_agenda}

\section{Conclusion}
The AI-native vision ultimately depends on a change in the human role: people should be able to govern intent, acceptable risk, and exceptional decisions without having to reproduce every engineering action that agents perform. Verification-first assurance is therefore not a peripheral quality activity; it is a condition for responsible delegation. Existing verification and validation techniques provide many of the necessary building blocks, but they do not yet provide assurance closure as an integrated, machine-operable capability.

This paper has framed that missing capability through six gaps, a corresponding high-level architecture, and four research questions. The central research challenge is to make the assurance reasoning around existing methods, what must be assured, what evidence is needed, whether the evidence should be believed, when it ceases to apply, and what authority it justifies, operate at the speed and scale of agentic development while preserving meaningful human control. Closing that gap is, in our view, necessary for the full AI-native large-scale agile software development vision to move from increasingly capable code generation to trustworthy, bounded delegation of R\&D authority.

\renewcommand*{\bibfont}{\scriptsize}
\printbibliography[title={References}]

\end{document}

%% file: sections/01_position.tex
\section{Introduction}
The AI-Native Manifesto proposes a form of large-scale agile software development in which humans progressively shift from implementation work to supervisory control, while remaining responsible for intent, risk, policy, and exceptional decisions, as agents assume increasing engineering authority \parencite{britto2026manifesto}. Its verification-first assurance principle is a prerequisite for that transition: delegation is meaningful only if the system can provide adequate grounds for trusting what agents produce.

We develop that unresolved requirement as \emph{assurance closure}: the ability to establish what must be true, determine and obtain appropriate evidence, judge the credibility of that evidence, preserve its validity as the system changes, and use the resulting uncertainty to bound agent authority. The individual techniques needed for these tasks already exist in many forms. The research challenge is to make the meta-assurance process that selects, connects, interprets, and maintains them sufficiently machine-operable for AI-native R\&D. This paper is therefore a position paper and research agenda rather than a proposal for a new verification method. 

The remainder of the paper is organized as follows: Section~2 reviews adjacent work and identifies six residual gaps that prevent assurance closure. Section~3 presents a high-level architecture whose six capabilities provide an integrated response to those gaps. Section~4 turns that architecture into four research questions that must be addressed before it can support dependable, human-on-the-loop delegation. Section~5 concludes by returning to the role of assurance in realizing the full AI-native vision.

%% file: sections/02_gaps.tex
\section{Related Work}
Assurance closure builds on several mature areas rather than replacing them. Requirements engineering and formal specification provide ways to make expected behavior more precise; recent agentic work shows that models can increasingly generate and reason about formal specifications, but also that accepted formalizations may still omit assumptions or admit incorrect behavior \parencite{hamblin2026specbench,agarwal2026verusspecgym}. Formal methods, static analysis, testing, and model-based techniques provide complementary forms of evidence. Industrial experience, such as Amazon's ShardStore, shows the value of decomposing correctness into properties and applying different techniques where they are most useful \parencite{bornholt2021shardstore}.

Other fields address how assurance evidence is organized and maintained. Assurance cases structure claims, evidence, and reasons for doubt \parencite{goodenough2012confidence}, while DARPA ARCOS targets greater automation in evidence evaluation and assurance-case construction \parencite{darpaArcos}. Digital-twin research emphasizes that simulation evidence is useful only when the model is credible for its intended purpose \parencite{shao2023credibility}. Continuous and runtime assurance, in turn, show how assurance arguments and safety constraints can remain active as systems operate and change \parencite{sljivo2024dynamic,slagel2024rta}. These foundations are substantial. What is still missing is an integrated mechanism that enables an agentic R\&D system to perform much of the assurance reasoning that expert engineers currently provide for them.

From this landscape, we identify six gaps, ordered as a stepwise transformation toward greater delegated authority.

\textbf{GAP 1---Autonomous specification adequacy.} Before an agent is allowed to implement a change, the system needs to know whether the available description of the intended behavior is good enough to act on. This is more than translating prose into a formal language. Requirements may be incomplete, mutually inconsistent, too weak to distinguish correct from incorrect behavior, or based on assumptions that have never been made explicit. Recent benchmarks show that specification-level reasoning remains difficult even when formal tools are available \parencite{hamblin2026specbench,agarwal2026verusspecgym}. The gap is therefore the ability to detect that the current specification is \emph{not yet adequate for delegation} and to identify what is missing before implementation proceeds.

\textbf{GAP 2---Assurance strategy synthesis.} Once the expected behavior is sufficiently clear, the next question is what evidence should justify a particular change. Different properties call for different techniques: a local invariant may be suited to formal proof, an API contract to static or property-based checks, a distributed failure scenario to simulation or fault injection, and an operational property to runtime monitoring. Risk-based V\&V already makes such choices in practice, and heterogeneous assurance has proven valuable in industry \parencite{bornholt2021shardstore,shaw2026assurance}. What remains largely human work is constructing a proportionate assurance plan for each change: deciding which claims matter, how strongly they must be supported, which evidence can be reused, and where human judgment is still required.

\textbf{GAP 3---Scalable assurance evidence generation.} An assurance plan is useful only if the required evidence can actually be produced at AI-native speed. Existing methods already provide many ways to generate evidence, including theorem proving, model checking, static analysis, testing, fuzzing, simulation, emulation, executable reference models, digital twins, and runtime monitors. The unresolved issue is how an agentic system can instantiate these methods at the \emph{right scope and fidelity} for the change at hand. For example, dynamic validation should not require reproducing the whole production system if only a small interaction is relevant. The system may instead need to construct a focused executable environment that contains only the services, state, workloads, faults, and assumptions needed to challenge the affected behavior. Whatever environment is used must itself be credible for the claim being evaluated \parencite{shao2023credibility}. The research problem is therefore scalable, change-focused generation of assurance evidence, not any single mechanism such as a digital twin.

\textbf{GAP 4---Evidence credibility and sufficiency.} Producing evidence is not the same as having a good reason to believe a claim. Assurance cases and defeater-based reasoning already provide ways to relate claims to evidence and to ask what reasons for doubt remain \parencite{goodenough2012confidence}. Agentic development adds a further complication: the code, specification, tests, simulations, and evaluations may share the same model family, retrieved context, or mistaken interpretation. Several apparently independent artifacts may therefore repeat the same error. The gap is the ability to assess whether a body of evidence is relevant, sufficiently strong, sufficiently independent, and free of unresolved defeaters for the decision being made. This should not be reduced prematurely to a single synthetic confidence score.

\textbf{GAP 5---Assurance validity under continuous change.} Assurance is never permanent. A proof may depend on an interface assumption that later changes; a test may no longer represent production behavior; a simulation model may become stale; or new operational evidence may invalidate a previously accepted claim. Continuous and dynamic assurance already recognizes this lifecycle problem \parencite{sljivo2024dynamic}. AI-native R\&D raises the required speed and granularity: each software or environment change should automatically reveal which claims, assumptions, models, tests, proofs, and observations are affected, which evidence remains reusable, and what must be regenerated. Without this capability, faster generation simply creates an assurance bottleneck downstream.

\textbf{GAP 6---Assurance-governed delegation and supervisory control.} Closing the preceding gaps still leaves a governance question: what should agents be allowed to do with the assurance available now? Runtime-assurance architectures already constrain autonomous components when safety conditions are threatened \parencite{slagel2024rta}. AI-native software development needs an analogous control loop over engineering authority. The ability to propose a change, edit code, merge, release, or deploy should depend on the current assurance state and the consequences of being wrong. At the same time, humans must retain enough visibility to understand unresolved uncertainty, intervene in exceptional cases, and reduce or revoke authority. The gap is therefore not simply about more automation, but about evidence-based, reversible delegation with meaningful human supervision.

%% file: sections/03_architecture.tex
\section{Assurance-Closure Architecture}
\begin{figure*}[t]
  \centering
  \includegraphics[width=0.99\textwidth]{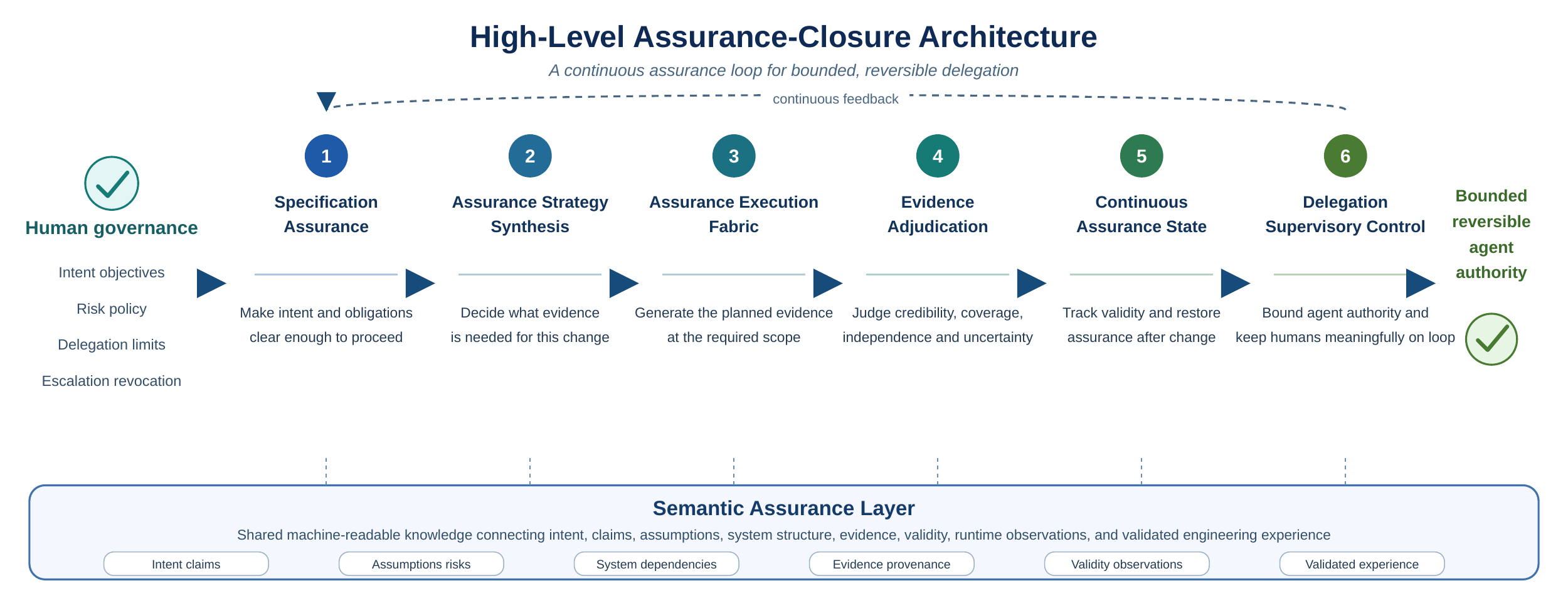}
  \caption{High-level architecture for assurance closure. Human governance defines intent, risk, policy, and limits on delegation. Six assurance capabilities progressively transform that input into bounded, reversible agent authority, while a shared semantic assurance layer maintains the knowledge needed across the loop.}
  \label{fig:architecture}
\end{figure*}

The six gaps suggest that assurance closure cannot be achieved by adding a single additional verification tool to the development pipeline. Figure~\ref{fig:architecture} presents a high-level architecture in which assurance becomes a continuous control function around agentic R\&D.

\textbf{C1--Specification Assurance} addresses GAP~1. It turns human intent, requirements, constraints, and known system behavior into machine-actionable obligations, but its key responsibility is to challenge them rather than merely formalize them. It should identify ambiguity, missing assumptions, conflicts, weak constraints, and important behavior that remains unspecified. Its output is therefore not ``a formal specification'' in a particular notation, but a specification state that is sufficiently complete and explicit for the next level of delegation, together with any unresolved questions that still require human input.

\textbf{C2--Assurance Strategy Synthesis} addresses GAP~2. Given the affected claims, architecture, risk, and uncertainty, it decides what evidence should be produced and at what depth. It may choose formal proof for some properties, testing or fuzzing for others, executable validation environments for interaction-heavy behavior, runtime monitoring for properties that cannot be fully established before deployment, or human review when judgment cannot be delegated. The capability is deliberately method-neutral: its purpose is to construct and continuously adapt an assurance strategy rather than to privilege a specific technology.

\textbf{C3--Assurance Execution Fabric} addresses GAP~3 by turning that strategy into evidence. It orchestrates heterogeneous tools and agents, scopes them to the affected parts of the system, and constructs the validation context required by the selected methods. This is where conventional static and formal techniques, BDD and model-based testing, property-based testing and fuzzing, simulation and emulation, targeted digital twins, runtime verification, and expert review can all participate. A central research concern is efficiency: the fabric should generate enough evidence to challenge the relevant claims without reproducing or re-verifying the entire system for every change.

\textbf{C4--Evidence Adjudication} addresses GAP~4. It interprets the evidence produced by C3 rather than treating tool success as proof of correctness. The capability must reason about relevance, coverage, provenance, method soundness, shared dependencies between evidence sources, uncertainty, and unresolved defeaters. It should also recognize when more evidence is needed or when apparently diverse evidence is actually based on the same assumption. The result is a structured assurance judgment that explains why the available evidence is or is not adequate for the decision under consideration.

\textbf{C5---Continuous Assurance State} addresses GAP~5. It maintains the dependencies between intent, claims, assumptions, system elements, assurance activities, evidence, and runtime observations. When something changes, C5 determines which assurances remain valid, which have become stale, and which parts of the assurance strategy must be rerun. The aim is incremental reassurance: preserve strong evidence when its assumptions still hold, while precisely invalidating the evidence that no longer applies.

\textbf{C6--Delegation and Supervisory Control} addresses GAP~6. It maps the current assurance state to concrete engineering permissions and escalation rules. An agent might be allowed to explore or prepare a change while being prevented from merging it; stronger evidence may permit merging but still require human approval for deployment. If evidence becomes invalid, authority should contract automatically. C6 also exposes the remaining uncertainty and the reasons for restricting access to humans, enabling them to supervise intent, risk, and exceptions rather than re-performing all implementation work.

A \textbf{semantic assurance layer} supports all six capabilities. It provides shared, machine-readable knowledge linking intent, claims, assumptions, risks, architecture, implementation, dependencies, assurance activities, evidence, provenance, defeaters, validity conditions, runtime observations, and validated engineering experience. Machine-readable assurance models and traceability are not new \parencite{foster2021sacm,darpaArcos}; the role of this layer is to provide the six capabilities with a common state on which automated assurance decisions can be made. It also preserves useful experience, such as failed assumptions, counterexamples, effective assurance strategies, and invalidated evidence, with explicit scope and provenance so that future assurance decisions can reuse knowledge without treating past experience as universally valid.

%% file: sections/04_agenda.tex
\section{Research Agenda}
The architecture in Figure~\ref{fig:architecture} is a hypothesis about the capabilities required for assurance closure, not an implemented solution. Realizing it requires advances in how those capabilities are automated, connected, and evaluated. We organize that work around four research questions.

\textbf{RQ1--Specification adequacy for delegation.} How can an agentic system determine that a specification is sufficiently complete, correct, consistent, and discriminating to justify a particular level of delegation? Research is needed to detect missing assumptions and weak constraints, generate counterexamples, compare independently derived interpretations, and decide when unresolved ambiguity requires human clarification rather than further autonomous work.

\textbf{RQ2--Automated assurance strategy and evidence generation.} Given a change, its affected claims, architecture, uncertainty, and consequence profile, how can the system construct a cost-effective portfolio of assurance activities and then instantiate it at the appropriate scope? This includes both selecting among proof, analysis, testing, fuzzing, simulation, targeted executable environments, runtime monitoring, and human review, and determining the minimum credible context in which each selected technique can provide useful evidence.

\textbf{RQ3--Evidence trustworthiness and sufficiency.} How can evidence produced partly by agents be assessed for applicability, soundness, provenance, dependence on shared assumptions, coverage, and unresolved defeaters? Agent-written tests, for example, are not automatically strong evidence; recent results show that increasing their volume may provide little improvement in repository-level issue resolution \parencite{chen2026agenttests}. Research should therefore focus on when heterogeneous evidence genuinely corroborates a claim and when it merely reproduces the same interpretation through different artifacts.

\textbf{RQ4--Continuous assurance and dynamic delegation.} How can assurance dependencies be maintained incrementally so that change automatically invalidates stale evidence, triggers proportionate re-assurance, and adjusts agent authority? Delegation should be reversible: weakened evidence or violated assumptions should reduce permissible actions, while restored evidence may expand them. This also requires studying the information humans need to supervise the loop effectively without becoming a manual verification bottleneck.

\textbf{Evaluation principle.} A useful demonstrator should expose the same system to changes that differ in affected claims, uncertainty, and risk. A meaningful result would be an assurance system that derives materially different and defensible strategies, constructs the required validation context, identifies remaining reasons for doubt, reuses evidence when justified, and grants different levels of authority. Evaluation should therefore consider not only task success but also false assurance, defect and defeater discovery, assurance cost, unnecessary human intervention, re-assurance latency after a change, and the calibration between the assurance state and delegated authority.